\documentstyle[12pt,lacoan]{article}
\begin{document}
\title{ 
THE CLEO-c PHYSICS PROGRAM
}
\author{
Thomas E. Coan \\
{\em Physics Department, Southern Methodist University, Dallas, TX 75275, USA} \\
{\em E-mail: coan@mail.physics.smu.edu}
}
\maketitle
\baselineskip=14.5pt
\begin{abstract}

The CLEO collaboration at the Cornell Electron Storage Ring (CESR) is
proposing a three-year experiment that specifically emphasizes charm and
QCD studies in the energy range $\sqrt{s}=3$--$5\,$GeV, utilizes the
existing detector and requires minimal modification of CESR to produce
a luminosity ${\cal{L}}=(1-5)\times10^{32}\,{\rm cm^{-2}sec^{-1}}$.  Key features
of this ``CLEO-c'' physics program are summarized.
\end{abstract}
\baselineskip=17pt
\newpage

\section{Introduction}

The impressive performance of the asymmetric B-factories at KEK and
SLAC in the $\Upsilon({\rm 4S})$ region and the large 
integrated luminosities ($\sim 400\,{\rm fb^{-1}}$) the corresponding
detectors, Belle and BaBar, are projected to collect in the next few
years renders additional contributions to B-meson physics by CLEO
problematic. Correspondingly, CLEO seeks to exploit its extensive
experience in heavy flavor physics and the excellent performance of
its nearly hermetic detector to perform focused studies of charm and
QCD physics in the energy range $\sqrt{s}=3-5\,$GeV. Indeed, CLEO
no longer runs at $\sqrt{s}\sim\Upsilon({\rm 4S})$ but began a year
long program in November 2001 to run at $\sqrt{s}=\Upsilon({\rm 1S}),
\Upsilon({\rm 2S}), \Upsilon({\rm 3S})$ and to collect
$\int{\cal{L}}dt=1-2\,{\rm fb^{-1}}$ at each resonance. This data
set is a factor of $10-20$ larger than the existing world data set of
similar type and will
provide $2-3\%$ measurements of leptonic widths and precise measurements
of a wide range of precision branching fractions to complement the
proposed QCD studies\cite{besson}.

The CLEO-c proposal\cite{cleoc} entails a 3-year run plan at ``charm production''
threshold: running at $\sqrt{s}=\Psi(3770)$ ($D\bar{D}$ threshold) in
2003 and collecting $\int{\cal{L}}dt= 3\,{\rm fb^{-1}}$; running in
2004 at $\sqrt{s}\sim 4100\,$MeV ($D_s\bar{D}_s$ threshold) and
collecting $\int{\cal{L}}dt= 3\,{\rm fb^{-1}}$; and running in 2005 at
$\sqrt{s}={\rm J}/\Psi(3100)$ and collecting $\int{\cal{L}}dt= 1\,{\rm
fb^{-1}}$. These integrated luminosities represent data sets that are
a factor of $15-500$ larger than corresponding data sets of BES~II and
MARK~III.

A focussed set of precision measurements of charm absolute branching
fractions and spectroscopy of $c\bar{c}$ quarkonia is the
core of the CLEO-c physics program.  The physics reach has been
carefully simulated using analysis tools and detector understanding
acquired by the CLEO collaboration from many years of heavy quark
studies.  Measurement of $D$ and $D_s$ leptonic decays permit
determinations of the decays constants $f_D$ and $f_{D_s}$,
respectively. Measurements of semileptonic decays will yield precision
determinations of semileptonic form factors 
and the CKM matrix elements $V_{cd}$ and
$V_{cs}$ and so provide a check of the unitarity of the matrix
itself. Semileptonic form factors can also be accurately
determined. Precise charm absolute branching fractions will normalize
important $B$ decay measurements and provide a robust means of
verifying and calibrating theoretical techniques of heavy quark
effective theory (HQET) and lattice gauge theory that can in turn be
applied to the $b$ sector to improve measurements of CKM matrix
elements $V_{ub},V_{cb},V_{td}$ and $V_{ts}$.

High statistics samples of $c\bar{c}$ data (as well as $b\bar{b}$ data
from the $\Psi({\rm nS})$ runs now underway) will permit efficient
searches for purely gluonic ``glueballs'' and quark-gluon ``hybrids''
in the mass range $1.5-2.5\,{\rm GeV/c^2}$. The existence of such
exotic states is an unambiguous prediction of QCD and the unambiguous
detection of such states is of fundamental importance. Determination
of the $J^{PC}$ quantum numbers through partial wave analysis, as well
as measurements of decay widths and branching fractions of putative
exotic states, will clarify the gluonic content of exotic
candidates. Additionally, CLEO-c will have sensitivity to physics
beyond the Standard Model by mounting searches for $D\bar{D}$-mixing,
$CP$ violation in the charm sector, and searches for a host of rare
$D$ and $\tau$ decays.

Minimal modifications to CESR are required to implement the lower
energy running in the range $\sqrt{s}=3-5\,$GeV. Supplemental
radiation damping will be induced to produce transverse cooling by the
insertion of 18 meters of $2\,$T wiggler magnets so that the
luminosity scales as ${\cal{L}}\sim s$ rather than the naive
${\cal{L}}\sim s^2$ expectation. The design luminosity is
${\cal{L}}=(1-5)\times10^{32}\,{\rm cm^{-2}s^{-1}}$, depending on
$\sqrt{s}$, with an anticipated beam energy uncertainty $\sigma_B \sim
1.2\,$MeV at $\sqrt{s}={\rm J}/\Psi$.  Prototype wiggler design and
construction are well advanced.

\section{Charm Threshold Running}

CLEO-c will not be able to compete against the asymmetric $B$ factory
detectors running at $\sqrt{s}=\Upsilon({\rm 4S})$ in the raw number
of charm events collected. However, running at charm threshold
production has distinct advantages over continuum $e^+e^-\rightarrow
c\bar{c}$ production. For example, the charged and neutral
multiplicities in $\Psi(3770)$ events are only 5.0 and 2.4,
respectively, reducing combinatorics and leading to high detection
efficiencies and low systematic errors. Additionally, charm events at
threshold are pure $D\bar{D}$, including the $\Psi(4140)$ decaying
into $D\bar{D}^{\ast}$, $D_s\bar{D_s}$ and $D_s\bar{D_s}^{\ast}$. No
additional particles from fragmentation are produced.

Low multiplicity events and pure $D\bar{D}$ states, coupled with the
relatively high branching fractions typical of $D$ decays, permit the
efficient implementation of ``double tag'' studies where one $D$ is
fully reconstructed and the other is studied in a bias free
fashion. This permits the determination of absolute branching
fractions with very low backgrounds. In leptonic and semileptonic
decays, the missing neutrino can be treated as a missing mass problem,
with a missing mass resolution better than one pion mass.
Finally, the quantum coherence of the $D\bar{D}$ states produced in
$\Psi(3770)\rightarrow D\bar{D}$ and $\Psi(4140)\rightarrow \gamma
D\bar{D}$ decays permit relatively simple techniques\cite{quantum_c} for measuring
$D\bar{D}$ mixing parameters and direct $CP$ violation.

\section{Absolute Branching Fractions}

\begin{figure}[htb]
\vspace{9.0cm} 
\includegraphics{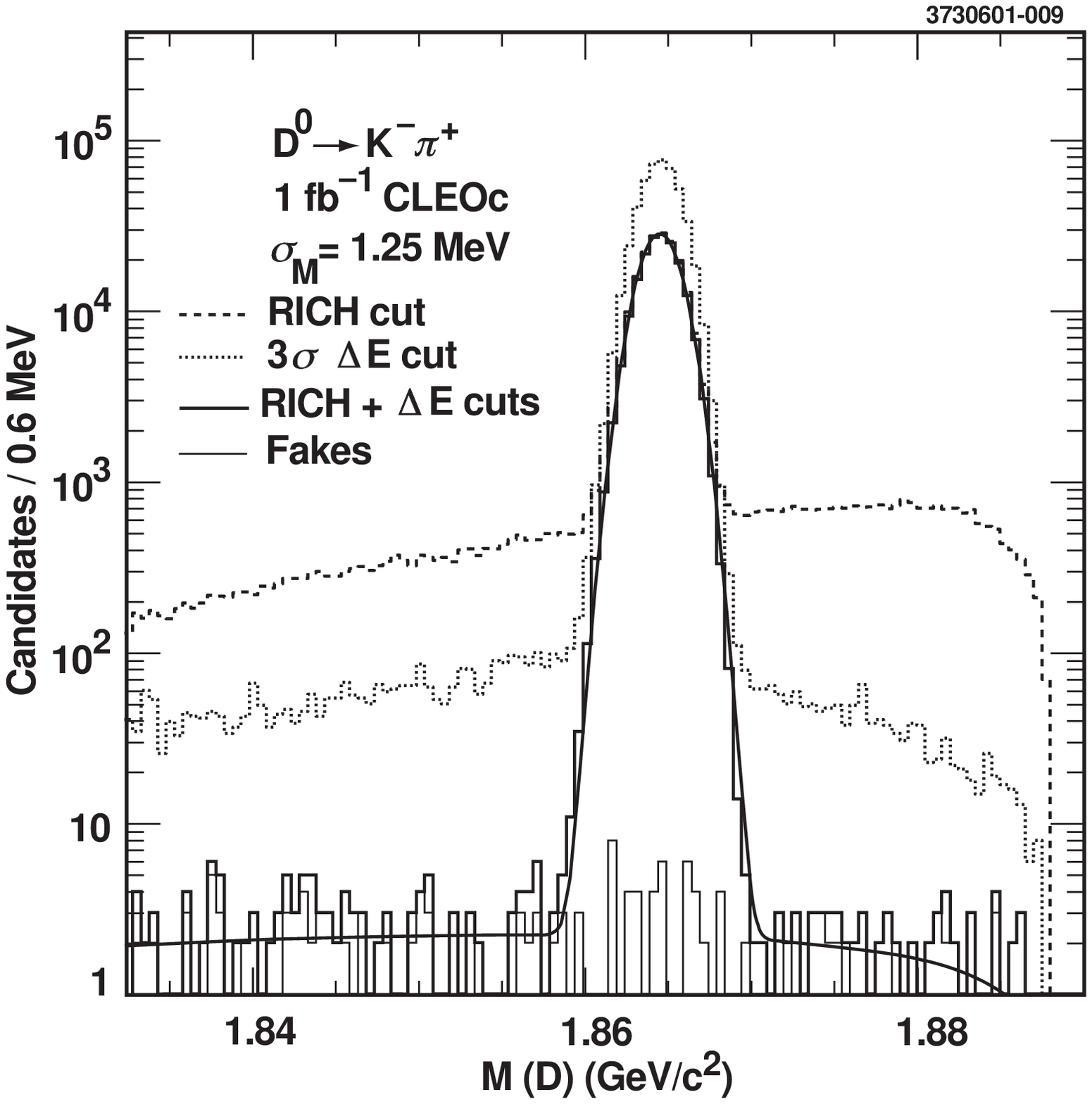} 

\includegraphics{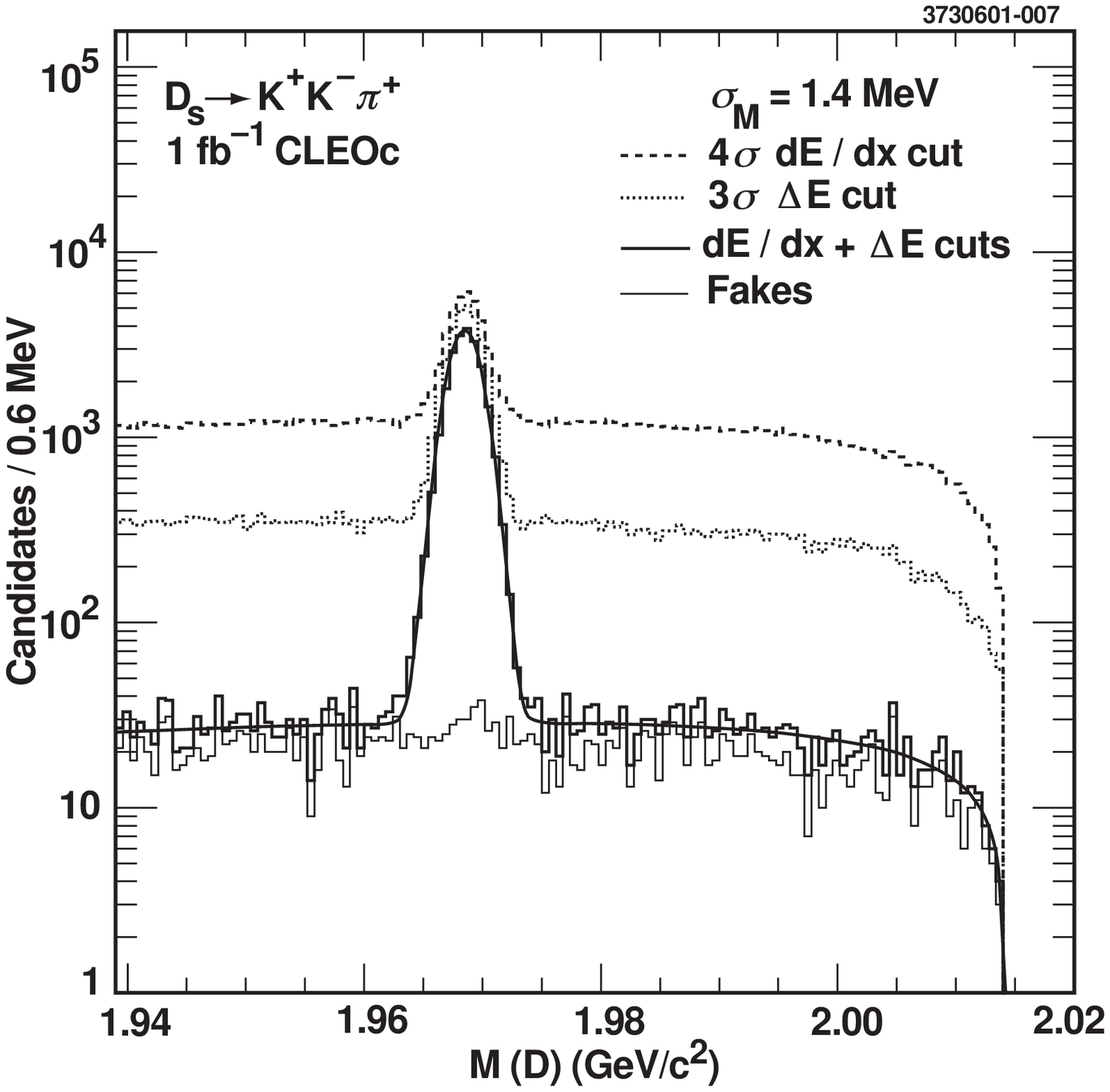} 
\caption{\it Reconstructed $D$ mass in
simulated $D^0\rightarrow K^-\pi^+$ decays (left)
and $D_s\rightarrow K^+K^-\pi^+$ decays (right) with CLEO-c.}
\label{tags}
\end{figure}

The combination of pure $D\bar{D}(D_s\bar{D_s})$ at
$\sqrt{s}=\Psi(3770)$ ($\sqrt{s}=4140\, {\rm MeV})$, typical charm
branching fractions of $(1-15)\%$, and a high reconstruction efficiency
for $D$-mesons, lead to a high net $D$-meson tagging efficiency of
$\sim 15\%$ for CLEO-c. Key selection criteria for $D$ candidates
include constraints on the energy difference between the D candidate
and the beam energy, the beam constrained mass of the $D$ candidate,
and particle identification cuts. Figure~\ref{tags} shows the
reconstructed $D^0$ mass for simulated $D^0\rightarrow K^-\pi^+$
decays (left) and the reconstructed $D_s$ mass for the $D_s
\rightarrow K^+K^-\pi^+$ mode. Only $1\,{\rm fb^{-1}}$ out of the
target $3\,{\rm fb^{-1}}$ of integrated luminosity is plotted and the
effects of sequential selection cuts are indicated. Note the
logarithmic vertical scale.

\begin{figure}[htb]
\vspace{9.0cm} 
\includegraphics{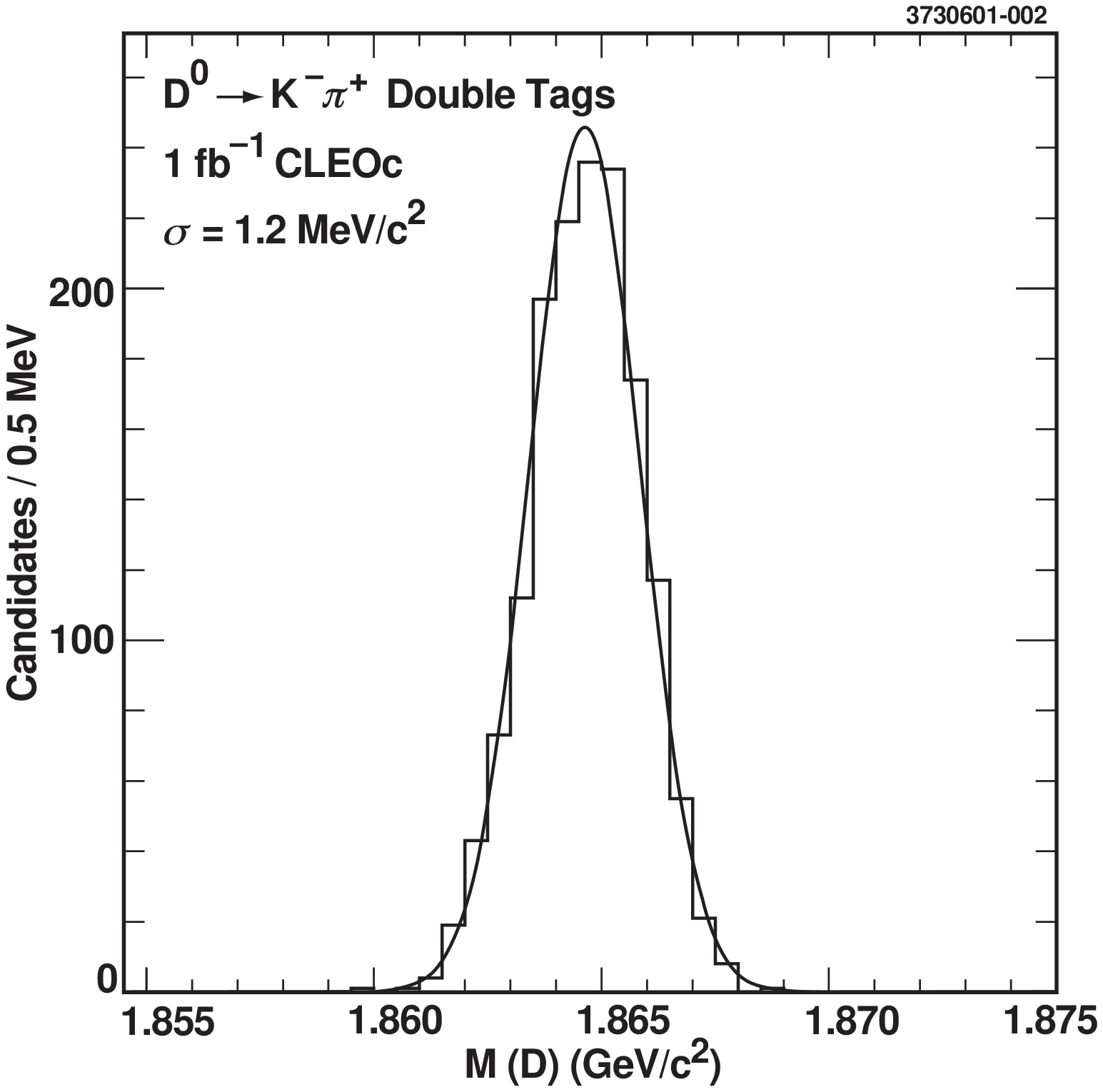} 

\includegraphics{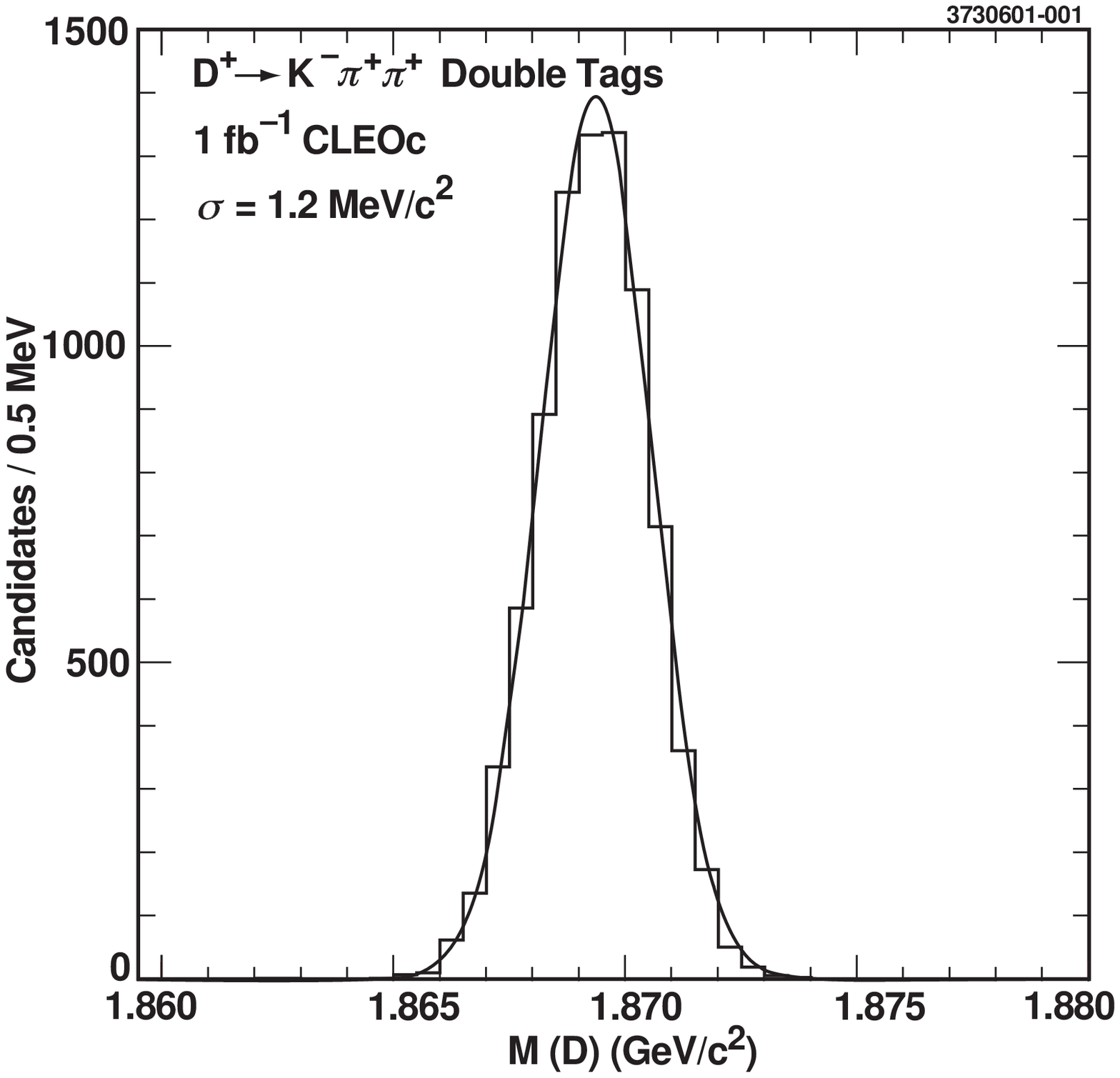} 

\caption{\it Reconstructed $D$ mass in
the $D^0\rightarrow K^-\pi^+$ decay mode (left) and
in $D^+\rightarrow K^-\pi^+\pi^+$ (right). A $D$ candidate
of the opposite flavor has also been found in
each event.}
\label{dbl}
\end{figure}

The technique for tagging a single $D$ candidate can of course be
extended to the second $D$-meson candidate in $e^+e^-\rightarrow
c\bar{c}$ threshold production events by essentially applying the
single-tag technique twice. Some additional constraints on the
flavor-defining charge of the second $D$ candidate, overall event
topology, and overall event energy and invariant mass are applied to
produce ``double-tag'' $D$ candidates.  Using a modified version of a
technique developed at MARK~III\cite{mark3}, CLEO-c can then precisely
measure absolute hadronic charm meson branching fractions using
double-tag events.  Figure~\ref{dbl} shows the reconstructed $D$ mass
from the $D^0\rightarrow K^-\pi^+$ mode (left) and from the
$D^+\rightarrow K^-\pi^+\pi^+$ mode.  In both cases the distribution
is made after another $D$ of the opposite flavor has been found in the
event and only $1\,{\rm fb^{-1}}$ of simulated data out of the
anticipated $3\,{\rm fb^{-1}}$ is used. Note the almost complete
absence of background in both distributions.  Table~\ref{tagtab}
compares the branching fraction precision for some key $D$ decay modes
anticipated with $3\,{\rm fb^{-1}}$ of CLEO-c data and the
corresponding PDG~2000 values.

\begin{table}[ht]
\centering
\caption{ \it Branching fraction precision of key $D$ decay modes projected
for $3\,fb^{-1}$ of CLEO-c data compared to PDG~2000 values.}
\vskip 0.1 in
\begin{tabular}{|l|c|c|c|} \hline
  \hskip 5.mm\rule[-8pt]{0pt}{22pt}  Mode  &  $\sqrt{s}$ (GeV) & $({\delta Br\over Br})_{PDG}$& $({\delta Br\over Br})_{CLEO-c}$ \\
\hline
\hline
 $D^0\rightarrow K^-\pi^+$   & $3770$ &$2.4\%$   & 0.6\% \\
 $D^+\rightarrow K^-\pi^+\pi^-$   & $3770$ &$7.2\%$   & 0.7\% \\
 $D_s\rightarrow \phi\pi$   & $4140$ & $25\%$   & 1.9\% \\
\hline
\end{tabular}
\label{tagtab}
\end{table}

\section{Leptonic and Semileptonic Decays}

Precision measurements of leptonic and semileptonic decays in the
charm sector are vital for determining CKM matrix elements that
describe the mixing of flavors and generations induced by the weak
interaction. The lowest order expression for the leptonic branching
fraction of a $D$-meson is given by\cite{rosner}

\begin{equation}
{\cal{B}}(D_q\rightarrow l\nu) = {G^2_F\over 8\pi}m_{D_q}m^2_l(1-{{m^2_l\over m^2_{D_q}}})
f^2_{D_q}|V_{cq}|^2\tau_{D_q},
\label{lepdcy} 
\end{equation}

\noindent where $f_{D_q}$ is the parameter that encapsulates the
strong physics of the process and $|V_{cq}|$ is the CKM matrix
parameter that encapsulates the weak physics and quantifies the
amplitude for quark mixing.  Measurements of leptonic branching
fractions can then be used to extract $f_{D_q}$ and, with additional
semileptonic measurements, $|V_{cq}|$. They will exploit fully tagged
$D^+$ and $D_s$ decays from running at the $\Psi(3770)$ and
$\sqrt{s}\sim 4140\,$MeV. $D_q\rightarrow \mu\nu$ decays are detected
in tagged events by observing a single track of the correct charge,
missing energy, and accounting for any residual energy in the
calorimeter.

\begin{figure}[htb]
\vspace{9.0cm} 
\includegraphics{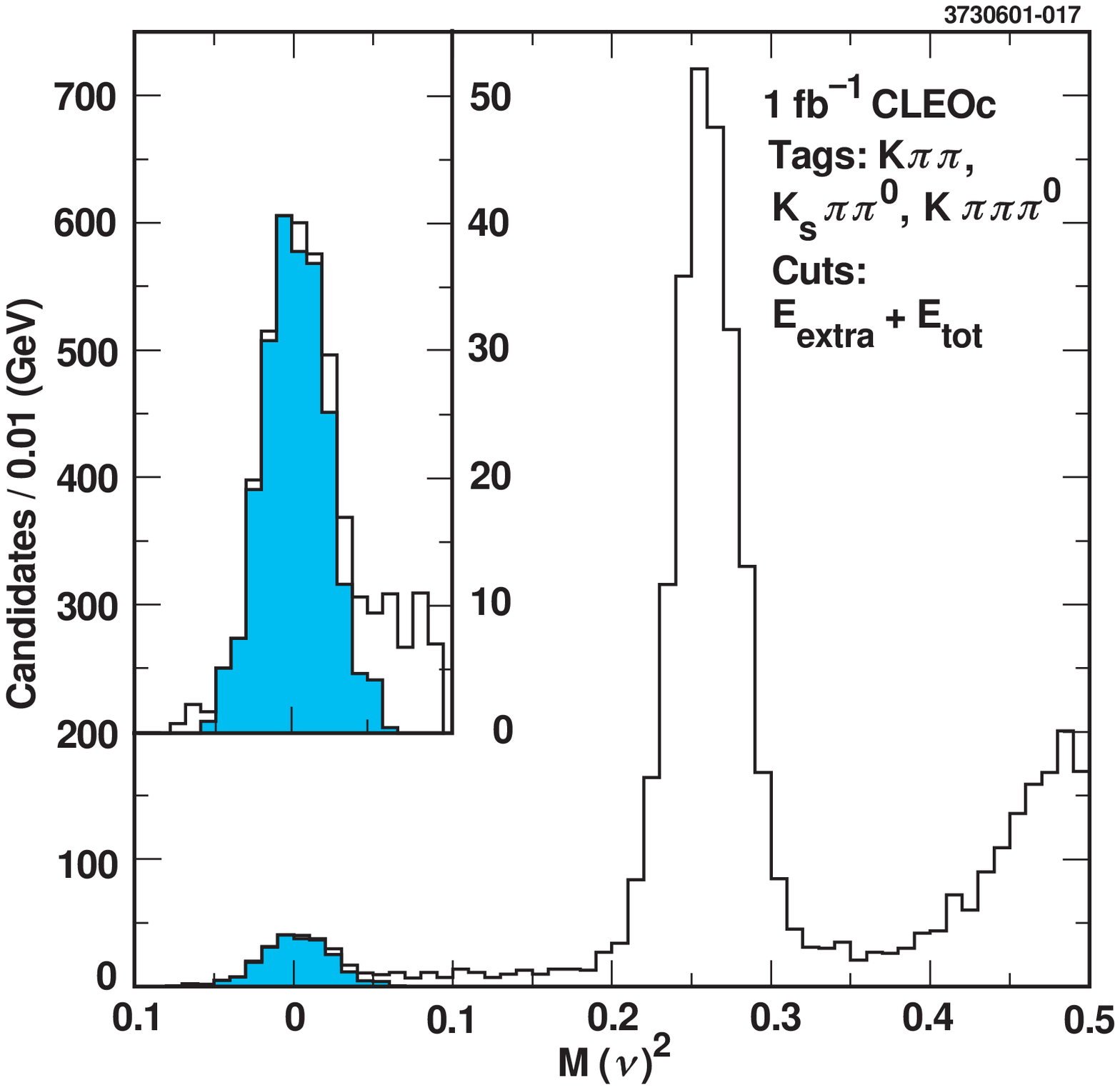} 

\includegraphics{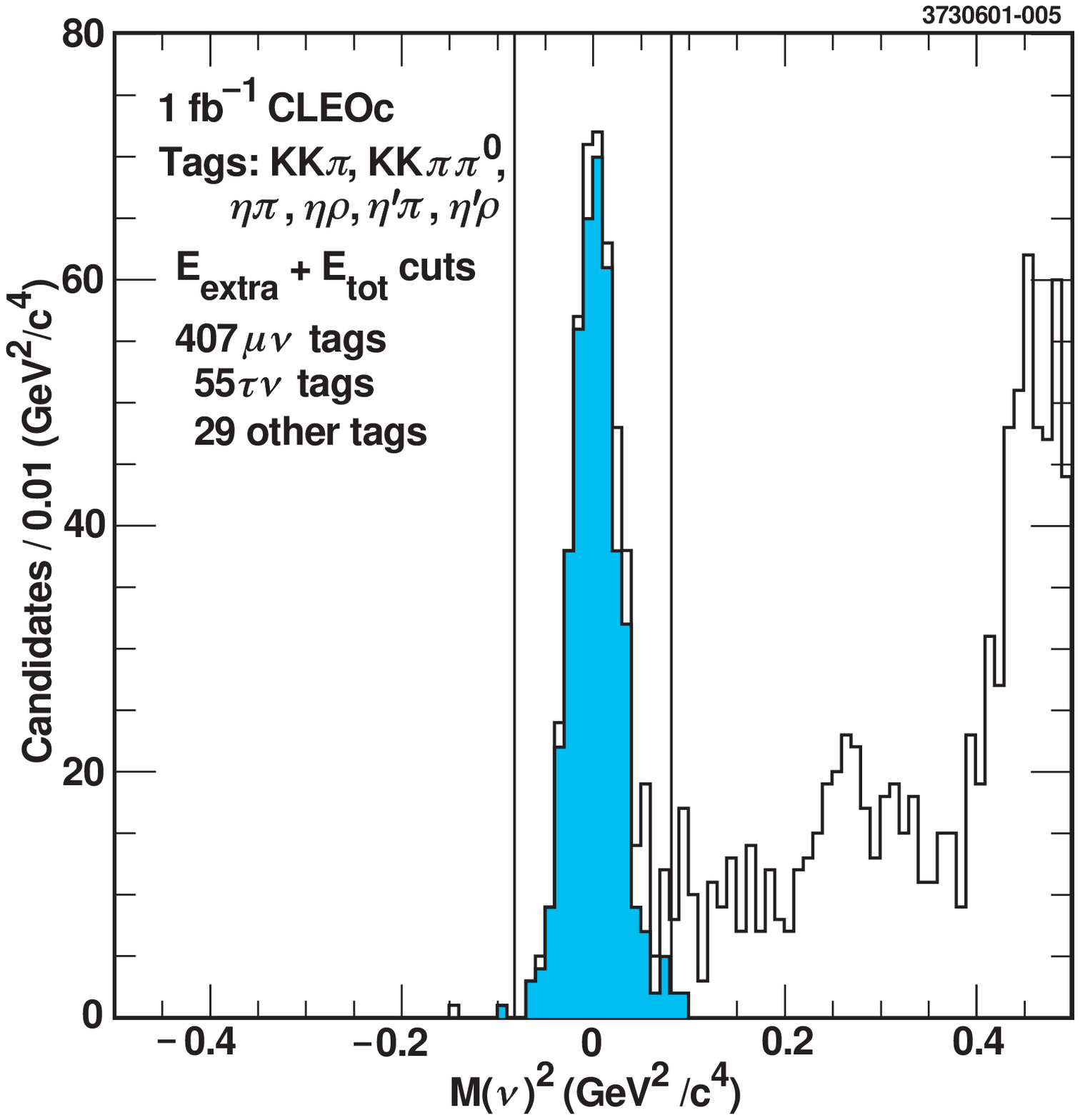} 

\caption{\it Reconstructed missing mass squared for simulated
$D^+\rightarrow \mu\nu$ candidates (left) and for $D_s\rightarrow
\mu\nu$ candidates (right). The shaded areas are the signal.}
\label{dmunu}
\end{figure}

Figure~\ref{dmunu} shows the reconstructed missing mass squared for
$D^+\rightarrow \mu\nu$ (left) and $D_s\rightarrow \mu\nu$
decays. Note the signal (shaded area) is cleanly separated from the
background. The decay $D_s\rightarrow \tau\nu$ is more complicated to
treat because of the secondary decay of the $\tau$ but still fully
accessible to CLEO-c. Table~\ref{fd} compares with PDG~2000 values the expected precision
for $D$-meson decay constants with $3\,{\rm fb^{-1}}$ of data and
assuming 3 generation unitarity.

\begin{table}[ht]
\centering
\caption{ \it Charm decay constant precision expected with
$3\,fb^{-1}$ of CLEO-c data compared to PDG~2000 values.}
\vskip 0.1 in
\begin{tabular}{|c|c|c|c|} \hline
   Decay Constant  &  \rule[-8pt]{0pt}{22pt}Mode (GeV) & $({\delta f\over f})_{PDG}$& $({\delta f\over f})_{CLEO-c}$ \\
\hline
\hline
 $f_{D_s}$   & $D^+_s \rightarrow \mu\nu$ & \rule{0pt}{14pt}$17\%$   & 1.9\% \\
 $f_{D_s}$  & $D^+_s \rightarrow \tau\nu$ & $33\%$   & 1.7\% \\
 $f_D$   & $D^+ \rightarrow \mu\nu$      & Upper limit   & 2.3\% \\
\hline
\end{tabular}
\label{fd}
\end{table}

The differential semileptonic decay rate for a $D$-meson to a
pseudoscalar $P$ is given by\cite{dplnu}

\begin{equation}
{d\Gamma(D\rightarrow Pl\nu) \over dq^2} = {G^2_F\over 24\pi^3}|V_{cq}|^2p^3_P|f(q^2)|^2,
\label{semilep}
\end{equation}
\noindent where the form factor $f(q^2)$ encapsulates the strong
physics.  Form factor measurements are a key means to test theory's
description of heavy quark decays. Precision measurements in inclusive
semileptonic decays can strenuously test heavy quark effective theory
(HQET)\cite{hqet} while exclusive decays are a rigorous testbed for
Lattice QCD (LQCD) calculations\cite{lqcd}. More progress in heavy
quark decay theory is necessary to extract precision values ($\delta
V/V\sim5\%$) of the CKM matrix elements $|V_{td}|$ and $|V_{ts}|$ from
data at present and planned $B$ physics experiments.
 
\begin{figure}[tb]
\vspace{9.0cm} 
\includegraphics{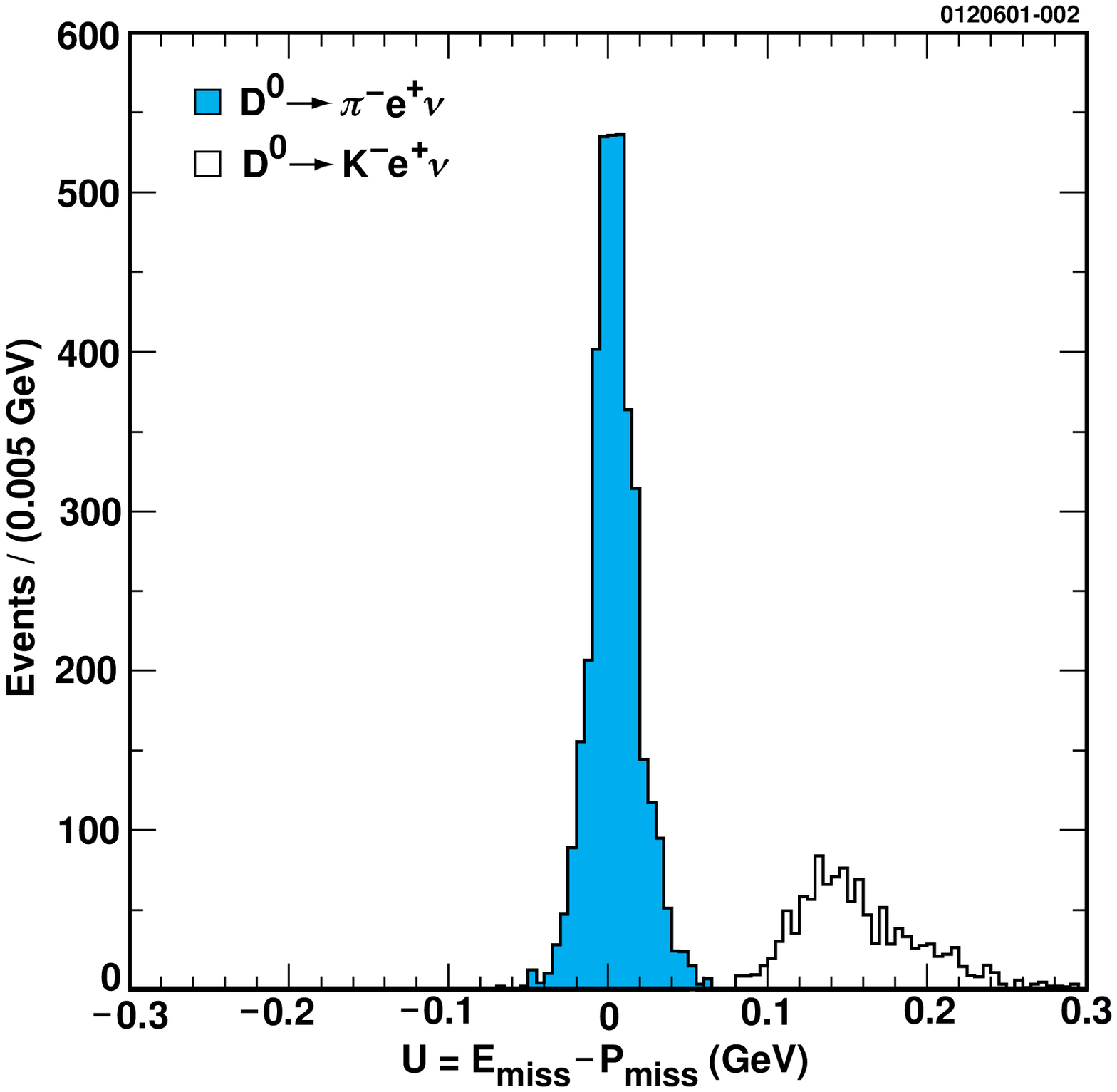} 

\includegraphics{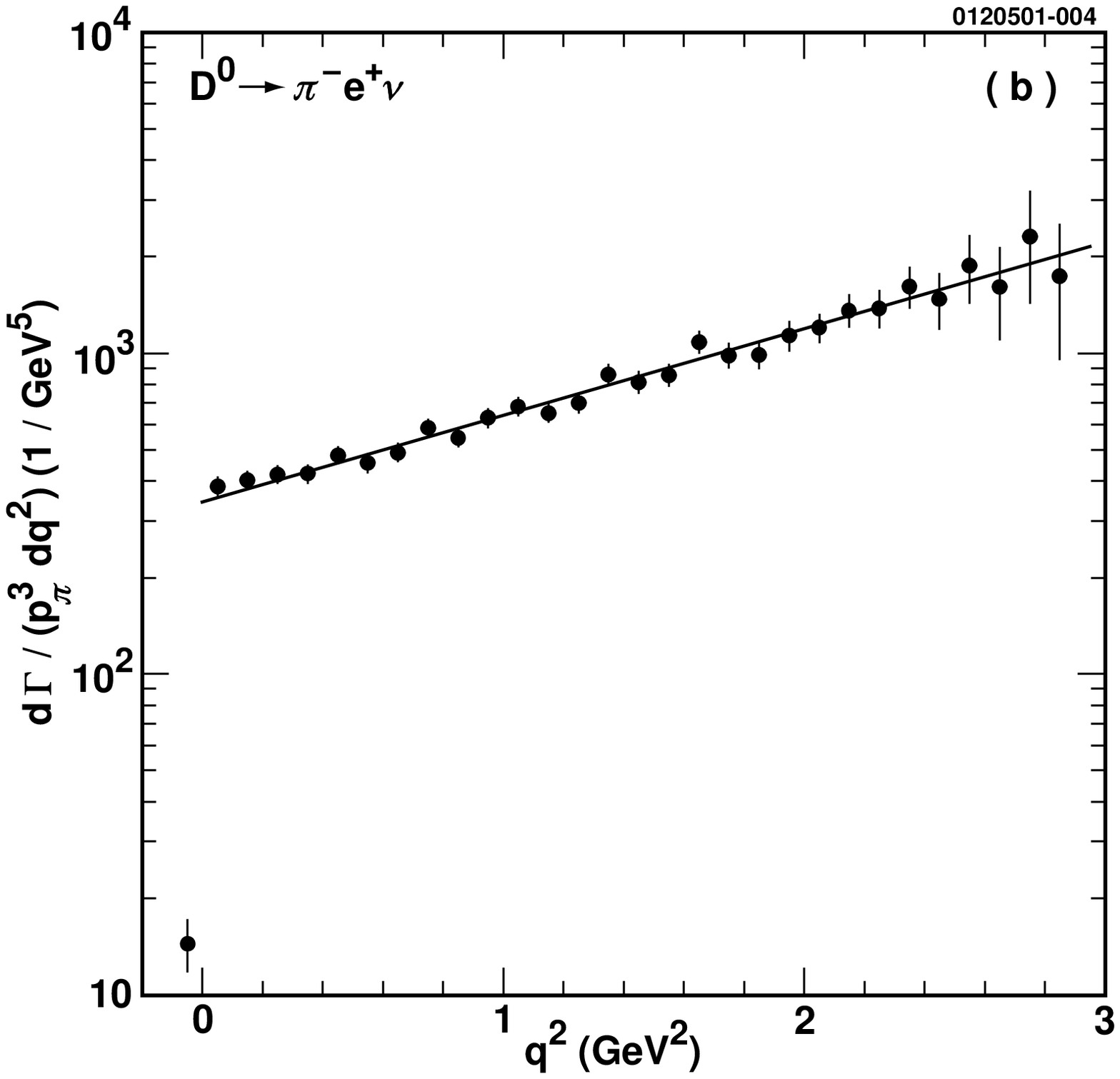} 

\caption{\it Reconstructed $D^0\rightarrow \pi^-e^+\nu$ decays (left)
as a function of the quantity $U$ described in the text. The signal is
the shaded area. A measurement of the form factor in the same decay
mode with a line fit to the data points (right).}
\label{slepdcy}
\end{figure}

Figure~\ref{slepdcy} shows on the left the reconstruction of the mode
$D^0\rightarrow \pi^-e^+\nu$ where the horizontal axis is the
difference $U$ between missing energy $E_{miss}$ and missing momentum
$P_{miss}$. The signal (shaded area) is well separated from the far
more abundantly produced $D^0\rightarrow K^-e^+\nu$ decays. The right
hand side shows a measurement of the form factor in the same
$D^0\rightarrow \pi^-e^+\nu$ decay and the line is a fit to the simulated
data using the parameterization $f=f(0)e^{\alpha q^2}$.
Table~\ref{sldcys} shows the expected precision in branching fraction
${\cal{B}}$ for some important semileptonic decays with CLEO-c for an
integrated luminosity of $3\,{\rm fb^{-1}}$ and the comparison with
PDG~2000 values.

\begin{table}[ht]
\centering
\caption{ \it Expected precision in the branching fraction
${\cal B}$ for important semileptonic decays with CLEO-c
and the comparison with PDG~2000 values.}
\vskip 0.1 in
\begin{tabular}{|l|c|c|} \hline
{\hfill \rule[-8pt]{0pt}{22pt}Mode \hfill} & $({\delta {\cal B}\over {\cal B}})_{PDG}$& $({\delta {\cal B}\over {\cal B}})_{CLEO-c}$ \\
\hline
\hline
 \ \ $D^0 \rightarrow K^-e^+\nu$ & \rule{0pt}{14pt}$5\%$   & 0.4\% \\
 \ \ $D^0 \rightarrow \pi^-e^+\nu$ & $16\%$   & 1.0\% \\
 \ \ $D^+ \rightarrow \pi^0e^+\nu$      & 48\% & 2.0\% \\
 \ \ $D_s \rightarrow \phi e^+\nu$      & 25\% & 3.1\% \\
\hline
\end{tabular}
\label{sldcys}
\end{table}

The absolute branching fraction for a semilptonic $D$ decay to a
pseudoscalar can be combined with a measurement of the $D$ lifetime $\tau_D$
to yield the total decay width:
\begin{equation}
\Gamma(D\rightarrow Pl\nu) = {{\cal{B}}(D^0\rightarrow Pl\nu)\over 
\tau_{D}}=\beta_{cq}V_{cq}, 
\label{ttldcy}
\end{equation}

\noindent with $\beta_{cq}$ given by theory.  Using
eqs.~\ref{lepdcy}, \ref{semilep} and \ref{ttldcy} and combining
measurements from leptonic and semileptonic decays make it possible
to measure charm decays constants directly, without the assumption of
3-generation unitarity, and to then determine the CKM matrix elements
$|V_{cd}|$ and $|V_{cs}|$, also without the unitarity assumption.
Table~\ref{vcq} shows the expected precision in $V_{cd}$ and $V_{cs}$
for CLEO-c with $3\,{\rm fb^{-1}}$ of integrated luminosity.

\begin{table}[ht]
\centering
\caption{ \it Expected precision in $V_{cq}$ matrix elements with
CLEO-c and the comparison to PDG~2000 values.}
\vskip 0.1 in
\begin{tabular}{|l|c|c|} \hline
{\hfill \rule[-8pt]{0pt}{22pt}${\delta V\over V}$\hfill} & CLEO-c & PDG~2000 \\
\hline
\hline
 \ \ ${\delta V_{cd}\over V_{cd}}$ & $1.6\%$   & \rule[-14pt]{0pt}{30pt}7\% \\
 \ \ ${\delta V_{cs}\over V_{cs}} $      & 1.7\% & \rule[-8pt]{0pt}{18pt}11\% \\
\hline
\end{tabular}
\label{vcq}
\end{table}

\section{QCD Probes}

CLEO-c will probe the low-energy nonperturbative structure of QCD with
new precision. QCD predicts the existence of bound hadronic states
in the mass range $\sim(1.5 -2.5)\,{\rm GeV/c^2}$ in
which gluons are both constituents and the source of the binding
force. Both fully gluonic ``glueballs'' and quark-gluon ``hybrids''
are novel forms of matter whose existence has yet to be unambiguously
demonstrated. Their detection and study will be a major focus of
$J/\Psi$ running.  With an expected CESR luminosity of
${\cal{L}}=2\times 10^{32}\,{\rm cm^2/sec^{-1}}$ at $\sqrt{s}=J/\Psi$,
CLEO-c expects to collect $1\times10^9$ $J/\Psi$ events.

Radiative $J/\Psi$ decays are a fruitful environment to search for
glue rich hadronic matter\cite{jrad} and CLEO-c will collect roughly
60 million $J/\Psi\rightarrow\gamma X$ decays with its projected
$1\,{\rm fb^{-1}}$ of integrated luminosity from $J/\Psi$ running. The
excellent energy resolution ($\sigma_{E}=4\,{\rm MeV}$ at $E_{\gamma}=
100\,{\rm MeV}$) and large solid angle coverage of CLEO's calorimeter
will permit efficient use of partial wave analyses to determine
absolute branching fraction for scalar resonances $X\rightarrow
\pi\pi, KK, \eta\eta, p\bar{p}$.

\begin{figure}[p]
\vspace{6.0cm} 
\includegraphics{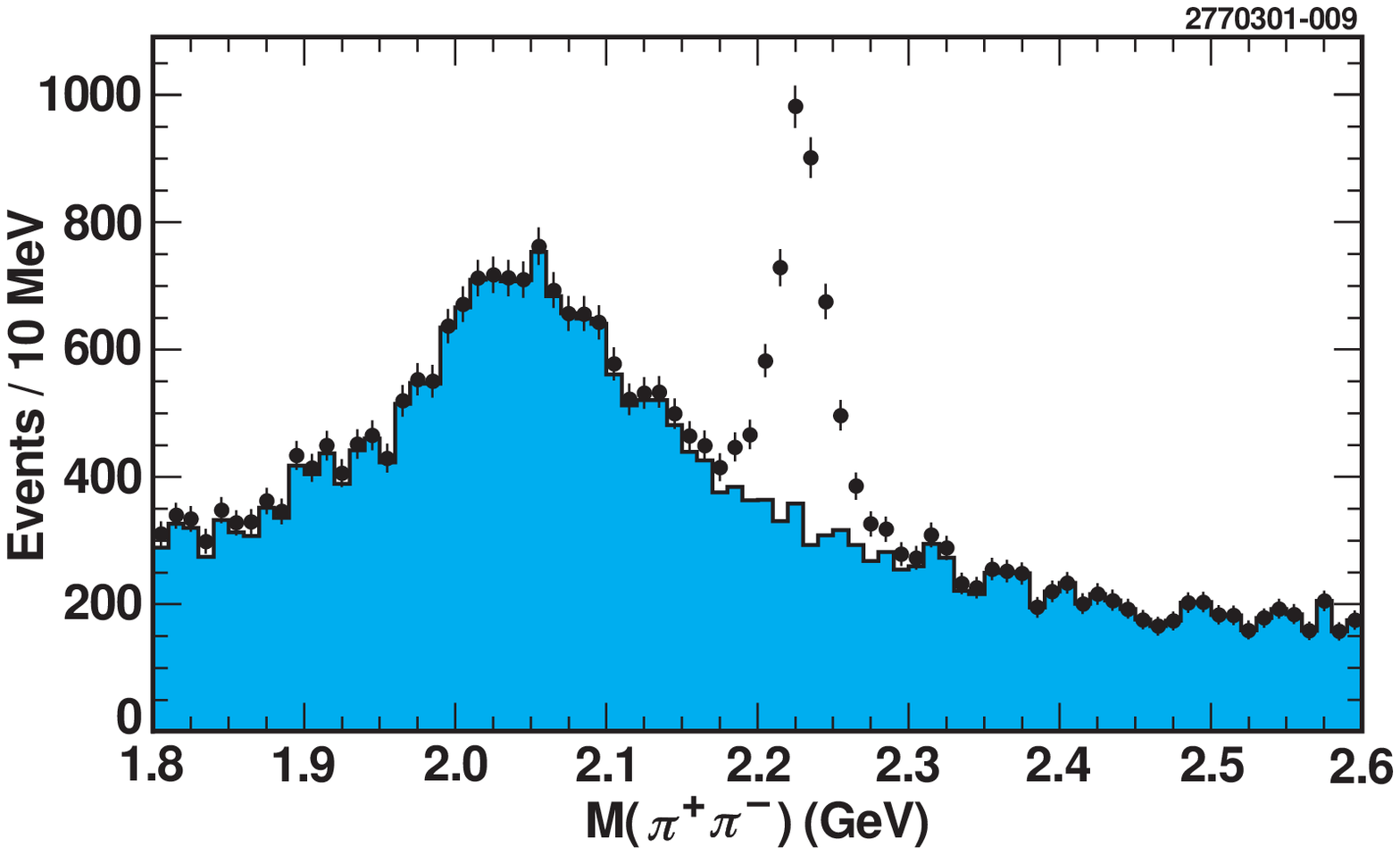} 

\includegraphics{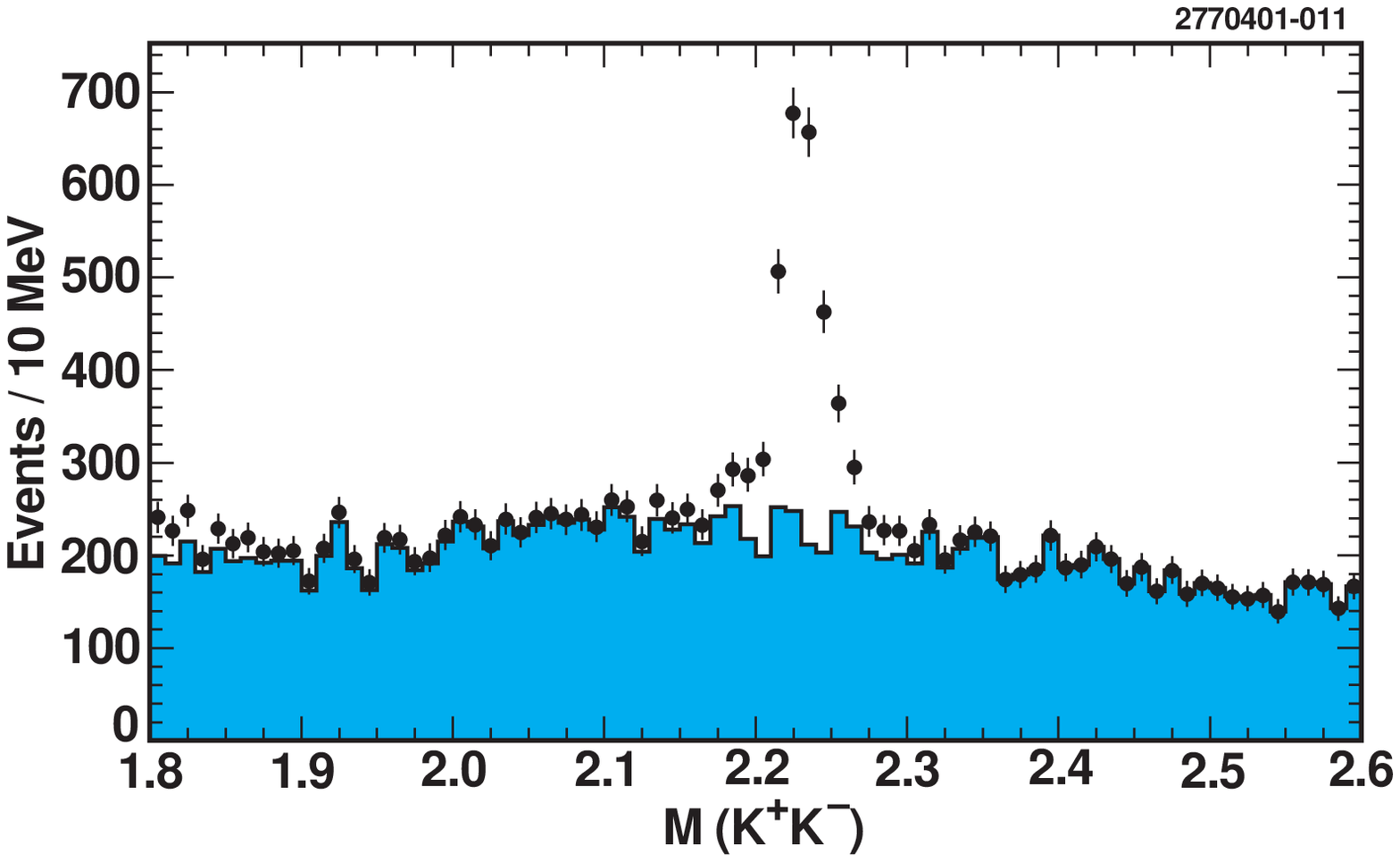} 

\caption{\it Expected invariant mass spectrum with CLEO-c
for $J/\Psi \rightarrow \gamma \pi^+\pi^-$ (left)
and $J/\Psi\rightarrow \gamma K^+K^-$ (right). The shaded portion in each plot
corresponds to hadronic and radiative backgrounds. The peaks are
attributed to the $f_J(2220)$.}
\label{fj}
\end{figure}

For example, with CLEO-c's projected data set and if the branching fraction
measurements from BES\cite{bes} are indeed correct for the $f_J(2220)$, then
CLEO-c will see many thousands of events in a variety of exclusive
$f_J(2220)$ decay modes. Figure~\ref{fj} shows the
invariant mass spectrum for $J/\Psi \rightarrow \gamma\pi^+\pi^-$
(left) and for $J/\Psi \rightarrow \gamma K^+K^-$.  Using only 1/6 of the
projected data set, the $f_J(2220)$ is clearly distinguishable from
the hadronic and radiative backgrounds indicated by the shaded
regions. Firmly establishing or debunking the existence of the
$f_J(2220)$ is a CLEO-c priority.

\begin{figure}[hb]
\vspace{6.0cm} 
\includegraphics{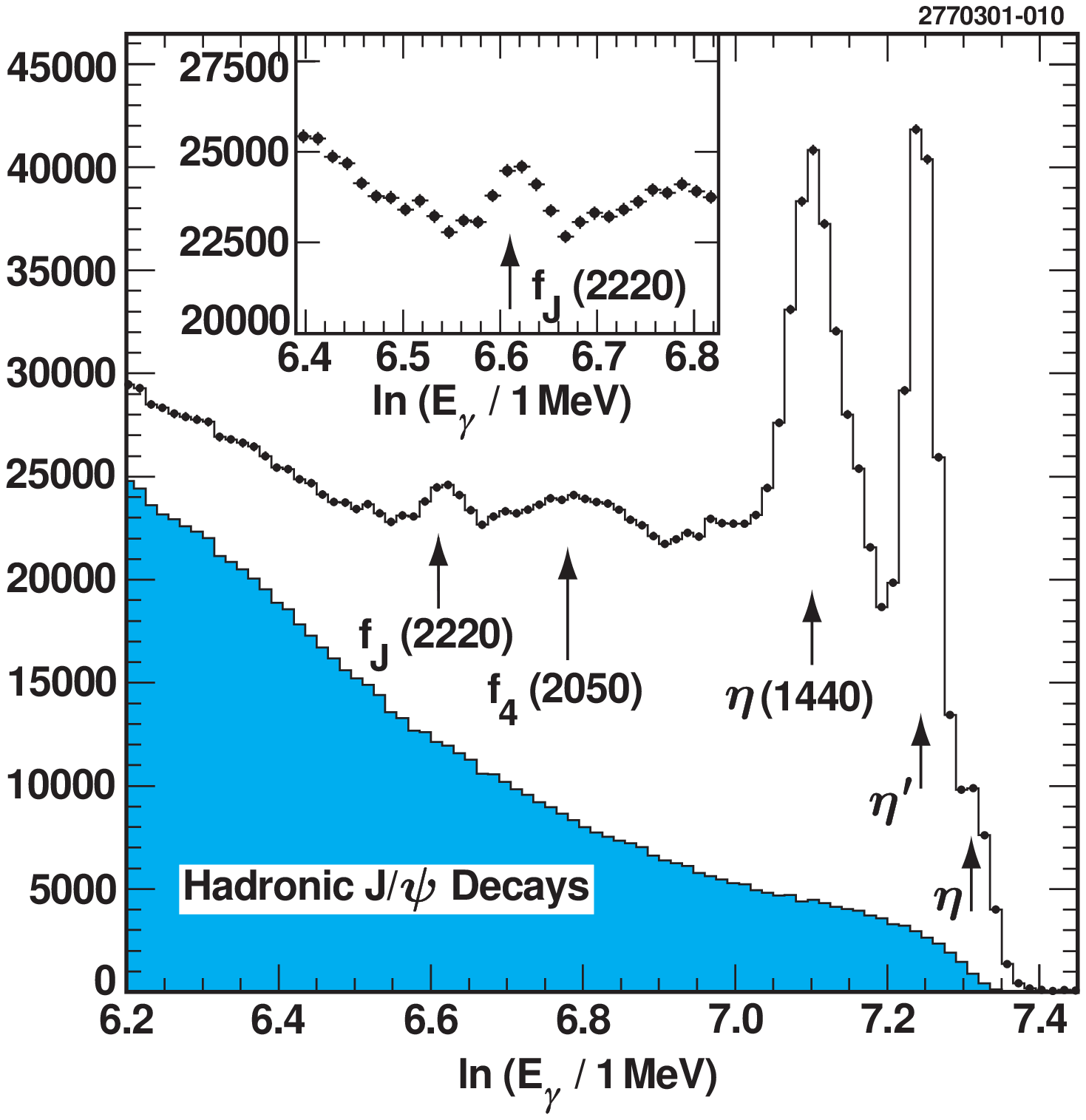} 

\caption{\it The inclusive photon spectrum for radiative
$J/\Psi$ decays with CLEO-c. The shaded region shows the
background from hadronic $J/\Psi$ decays.}
\label{jincl}
\end{figure}

The inclusive photon spectrum from radiative $J/\Psi$ decays is also a
powerful means to search for new glue-rich hadronic states.  As an
example, Figure~\ref{jincl} shows such an inclusive photon spectrum
for only 60 million $J/\Psi$ decays under the assumption
${\cal{B}}\rightarrow \gamma f_J(2220) = 8 \times 10^{-4}$. The
$f_J(2220)$ peak is clearly visible from the hadronic $J/\Psi$
background shown as the shaded region.  Due to its nearly hermetic
structure (93\% of $4\pi$), the CLEO-c detector is highly efficient at
rejecting events of the type $J/\Psi\rightarrow \pi^0X$ where one of
the photons from the $\pi^0$ gets lost. With its projected data set of
$10^9$ $J/\Psi$ events, CLEO-c should be able to detect any narrow
resonances in radiative $J/\Psi$ decays with a branching fraction of
${\cal{O}}(10^{-4})$ or larger. An exisiting data set of $25\,{\rm
fb^{-1}}$ of $\gamma\gamma$ events will be a useful crosscheck for
putative glue-rich states.

\section{Detector Performance}
The performance of the current CLEO detector is well-suited to the
CLEO-c physics program. In some sense the detector is overdesigned
since it was designed for running at $\sqrt{s}=\Upsilon(4S)$. Both the
tracking system and the calorimeter cover $93\%$ of $4\pi$ steradians,
providing the near hermeticity vital for partial wave analyses and for
efficient background rejection.  The central drift chamber has a
fractional momentum resolution $\sigma_p/p=0.35\%$ at $p=1\,{\rm
GeV/c}$ and a $D\rightarrow K\pi$ mass resolution $\sigma\sim
6.3\,$MeV. The CsI calorimeter has excellent energy resolution,
$\sigma_e/E = 2.2\%$ at $E=1\,$GeV and $5-7\,$MeV $\pi^0$ mass
resolution. The dedicated charged hadron identification system, a ring
imaging Cherenkov detector that covers 83\% of the full solid angle
that is used in conjunction with $dE/dx$ measurements from the drift
chamber, provides $10-200$ sigma $K/\pi$ separation from $470\,$MeV
up to the kinematic limit for $D$ decay daughters at CLEO-c collision
energies.

\section{Summary}
CLEO-c is a timely proposed experiment designed to make important
measurements in charm physics and QCD with projected data sets many
orders of magnitude larger than current world sets of comparable
type. These measurements will provide important measurements of CKM
matrix elements and strenuously test theory's description of heavy
quark decay as well as QCD's prediction of exotic gluonic states of matter.
Table~\ref{bfac} summarizes some important CLEO-c measurements,
projecting $3\,{\rm fb^{-1}}$ of integrated luminosity, and compares
them to the expected reach of the current asymmetric B-factories,
assuming $400\,{\rm fb^{-1}}$ of integrated luminosity. The relative
advantages of charm threshold running lead to favorable overall
errors.  The current CLEO detector is fully capable of providing
measurements with the appropriate resolution in particle energy,
tracking and charged hadron identification to accomplish CLEO-c
physics goals.

\begin{table}[ht]
\centering
\caption{ \it Comparison of the physics reach for some important
measurements between CLEO-c and the asymmetric B-factories and
PDG~2000 values.}
\vskip 0.1 in
\begin{tabular}{|l|c|l|c|} \hline
& CLEO-c & $B$-factories & PDG \\
\hline
\hline
 \ \ $f_D$ & $2.3\%$   & 10--20 \%   & -- \\
 \ \ $f_{D_s}$  & 1.7\% & 6--9\%      & 19\%  \\
 \ \ ${\cal{B}}(D^+\rightarrow K\pi\pi)$  & 0.7\% & 3--5\%      & 7\%  \\
 \ \ ${\cal{B}}(D_s\rightarrow \phi\pi)$  & 1.9\% & 5--10\%      & 25\%  \\
 \ \ ${\cal{B}}(D_s\rightarrow K\pi)$    & 0.6\% & 2--3\%      & 2\%  \\
\hline
\end{tabular}
\label{bfac}
\end{table}

\section{Acknowledgements}
I would like to thank the conference organizers for both the
invitation and the conference's excellent organization. The assistance
of I. Shipsey in the preparation of this talk is appreciated. The
author's work is supported by the U.S. Dept. of Energy under contract
DE-FG03-95ER40908.

\end{document}